\documentclass[prl,superscriptaddress,twocolumn]{revtex4}
\usepackage{graphics}

\begin{document}

\newcommand{\be}{\begin{equation}}
\newcommand{\ee}{\end{equation}}
\newcommand{\bea}{\begin{eqnarray}}
\newcommand{\eea}{\end{eqnarray}}
\def\vec#1{{\mathbf{#1}}}

\newcommand{\note}[1]{\textsc{NOTE: #1}}

\newcommand{\unit}[1]{\;{\rm #1}}

\newcommand{\kgd}{\unit{kg}\unit{d}}

\newcommand{\vearth}[1]{v_{\oplus,#1}}
\newcommand{\vex}{\vearth{x}}
\newcommand{\vey}{\vearth{y}}
\newcommand{\vez}{\vearth{z}}
\newcommand{\kmps}{\unit{km/s}}
\newcommand{\kpc}{\unit{kpc}}
\newcommand{\GeV}{\unit{GeV}}
\newcommand{\keV}{\unit{keV}}
\newcommand{\days}{\unit{days}}

\title{Comparing WIMP Interaction Rate Detectors with Annual Modulation
  Detectors}

\author{Craig J. Copi}
\email{cjc5@po.cwru.edu}
\affiliation{Department of Physics}
\author{Lawrence M. Krauss}
\email{lmk9@po.cwru.edu}
\affiliation{Department of Physics}
\affiliation{Department of Astronomy \\
Case Western Reserve University, 
10900 Euclid Ave., Cleveland OH 44106-7079}

\date{\today}

\begin{abstract}
  
  We compare the sensitivity of WIMP detection via direct separation of
  possible signal vs.\ background to WIMP detection via detection of an
  annual modulation, in which signal and background cannot be separated on
  an event-by-event basis.  In order to determine how the constraints from
  the two different types of experiments might be combined an adequate
  incorporation of uncertainties due to galactic halo models must be
  made.  This issue is particularly timely in light of recent direct
  detection limits from Edelweiss and CDMS, which we now demonstrate cannot
  be made consistent with a recent claimed DAMA annual modulation
  observation, even by including halo uncertainties.  On the other hand, we
  demonstrate that a combination of these two techniques, in the event of
  any positive direct detection signal, could ultimately allow significant
  constraints on anisotropic halo models even without directional
  sensitivity in these detectors.

\end{abstract}


\makeatletter
\@booleantrue\preprint@sw
\makeatother

\preprint{CWRU-P12-02}

\maketitle

The recent results from the WIMP direct detection experiments
CDMS~\cite{cdms} and the annual modulation-sensitive DAMA
detector~\cite{dama} appear inconsistent.  Indeed, they have been claimed
to be incompatible at the $>99.98\%$ level~\cite{cdms}.  This compatibility
estimate is based on the assumption that the dark matter halo in our
galaxy is well described by an isothermal sphere with a velocity dispersion
of $220\kmps$ (the standard model) for the the local velocity distribution
of WIMPs.  However, because these two different types of experiments are
searching for different WIMP signals the WIMP halo distribution can play a
key role in determining the ultimate constraints that one might derive from
the experimental results.  As a result, the actual level of inconsistency
between CDMS and DAMA is likely to be strongly model-dependent.  More
recently, however, a new result from the Edelweiss~\cite{edel} detector puts
a stronger bound on WIMP cross sections for much of the mass range that is
apparently favored by the DAMA result.  This makes the question of the
possible significance of halo model uncertainties more timely---namely, in
light of this new result, is there any room left for astrophysical
uncertainties to allow a reconciliation of the DAMA result with the
CDMS/Edelweiss?  Motivated by this fact, we explore here a wide range of
analytic halo models to examine their effects on the expected WIMP
signatures in these two different types of detectors, and present several
new ways of comparing the data.  We derive two main results: (1) Halo
model uncertainties do not allow a reconciliation of the DAMA result with
the CDMS/Edelweiss results, and (2) a combination of these two techniques
could ultimately provide sensitivity to anisotropies in a galactic WIMP
halo even if direct detection experiments do not have directional
sensitivity.

CDMS is a cryogenic detector that is capable of measuring the full energy
of the recoiling nucleus.  Since both the electronic and photonic channels
they have excellent background rejection and can thus look for individual
nuclear recoil events.  The signal they are searching for is an excess of
nuclear recoil events above their expected backgrounds.  With $10.6\kgd$ of
data they found no excess events above the expected neutron
background~\cite{cdms}.  Edelweiss uses the same detector technology as
CDMS but is in a deep underground site so that sophisticated background
subtraction is unnecessary.  With $7.4\kgd$ of data no WIMP events were found.
The two results are consistent and complementary.  CDMS is most sensitive
for low mass WIMPs ($m_\chi\le 35\GeV$) and Edelweiss is most sensitive for
higher mass WIMPS ($m_\chi\ge 35\GeV$).  When combined these two
experiments rule out essentially all of the DAMA region for the standard halo model.

DAMA consists of high purity NaI crystals run in a deep site.  They have no
particle identification and thus have no background subtraction
capabilities.  However, the detector mass is large so many events
(background and possible signal) can be recorded.  A modulation of the rate
is expected due to the Earth's motion around the Sun (and thus through the
WIMP halo) and any such modulation in the event rate provides a potential
WIMP signature.  With $57986\kgd$ of data they indeed report a modulation
signal that is claimed to be consistent with WIMP scattering.  From this a
mass and cross section for WIMPs can be determined.

Heuristically one can see that the halo model will affect the modulation
signal and the overall rate in different ways.  Consider the standard halo
model.  If we decrease the velocity dispersion (narrow the Gaussian) we
decrease the number of WIMPs in the tails of the distribution; in
particular we decrease the number of WIMPs with high velocity that can lead
to high energy nuclear recoils.  Thus the overall rate one expects to
observe is decreased.  However, at the same time the the size of the
modulation signal is increased because with fewer high velocity WIMPs the
Earth's motion around the sun becomes a larger perturbation on the net WIMP
velocity as measured in the laboratory frame.

Naturally these statements depend on the a number of factors.  In
particular the size of the effect depends on the masses of the WIMP and
target nucleus and on the velocity of the Earth (magnitude and direction)
with respect to the WIMP halo.  Here we will analyze a set of analytic halo
models. The procedure for calculating WIMP scattering rates is well
known~\cite{LS,jungman,chk}.  We briefly discuss some aspects here.  A more
detailed description of the analysis is given elsewhere~\cite{ckprep}.

The differential scattering rate for a WIMP of mass $m_\chi$ from a nucleus
with atomic number $A_N$ is given by
\be
  \frac{dR}{dQ} = \frac{\left(m_n+m_\chi\right)^2 A_N^2 \sigma_p \rho_0}{2m_n^2
      m_\chi^3} F^2 (Q) \int \frac{f (\vec v)}{|\vec v|} d^3 v.
\ee
Here $Q$ is the recoil energy of the nucleus, $m_n$ is the mass of a
nucleon (taken to be a proton), $F^2 (Q)$ is the nuclear form factor which
we take to be the standard Helm form~\cite{helm}, $\rho_0$ is the local
halo density of WIMPs, and $f (\vec v)$ is the local WIMP halo velocity
distribution.

For CDMS/Edelweiss the target nucleus is Ge ($A_{\rm Ge}=73$), the full
energy is measured so quenching is not an issue, and they are sensitive to
$10\textrm{--}100\keV$ recoils(CDMS) and $\ge20\keV$ (Edelweiss).  For DAMA
the detector consists of two nuclei, sodium ($A_{\rm Na}=23$) and iodine
($A_{\rm I}=127$).  Only the ionization energy is measured so only a
fraction of the energy is detected.  We use $q_{\rm Na}=0.30$ and $q_{\rm
  I}=0.09$.  Note that DAMA claims to observe a modulation signal for
measured recoil energies in the range $2\textrm{--}6\keV$.  We incorporate
their finite energy resolution in our calculations~\cite{damaexp}.

We probe a range of halo models ranging from spherically symmetric to
triaxial to discontinuous.  These models have been studied in the context
of the angular signal expected in future detectors possessing angular
resolution and more details can be found there~\cite{ck}.  Briefly, we
consider the isothermal model, an axisymmetric Evan's model, a triaxial
halo model, and a model of caustics that leads to WIMP flows in velocity
space.  For the isothermal model we consider dispersions of $v_0=170\kmps$,
$220\kmps$, and $270\kmps$ to account for observational uncertainties.
These models have been discussed in the context of WIMP detection.  For
demonstration purposes we explore here the effect of varying halo models on
the modulation signal in NaI detectors using parameters from DAMA and the
overall rate using parameters from CDMS\@.

\begin{figure*}
  \resizebox{5in}{!}{\rotatebox{-90}{\includegraphics{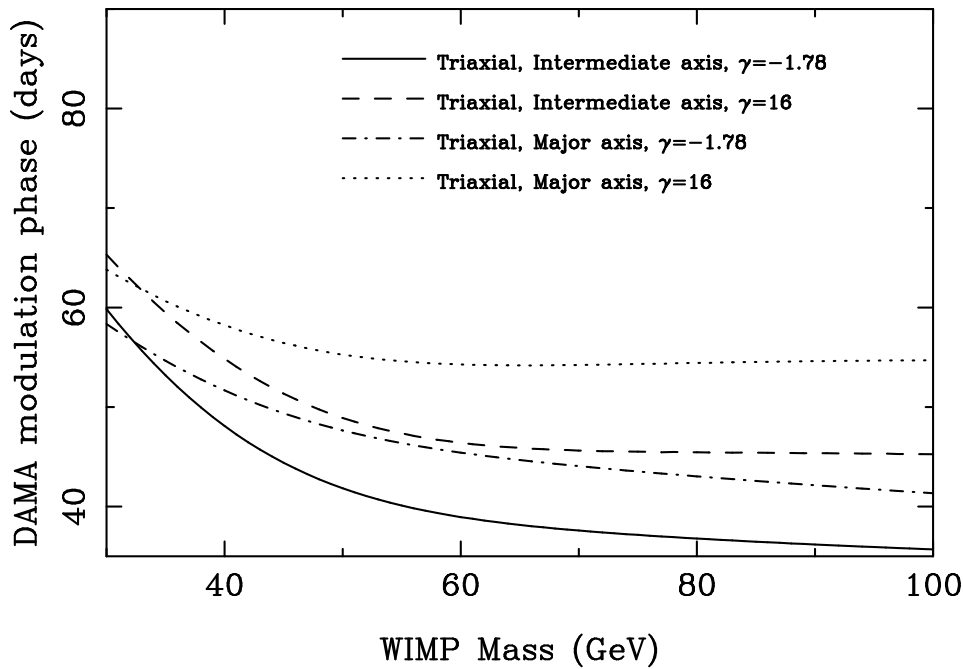}}}
  \caption{Modulation phase for non-symmetric halo distributions.  For the
    standard model the phase is $152.5\days$.}
  \label{fig:phase}
\end{figure*}

For the modulation signal there are two important parameters, the amplitude
of the modulation (related to the WIMP cross section) and the phase of the
modulation.  For symmetric models (such as the isothermal and Evans halos)
since all directions through the halo are approximately equivalent, the
modulation should be in phase with the motion of the Earth around the Sun.
Thus $t_p=152.5\days$.  For non-symmetric halos the direction of motion
through the halo is important and can lead to maximal scattering at
different times in the Earth's orbit.  The resulting phase for such models
is given in figure~\ref{fig:phase} as a function of WIMP mass.  Notice that
in all cases the phase is quite different from the standard model
$t_p=152.5\days$.

In comparing the CDMS/Edelweiss and DAMA results using the standard
isothermal halo, we confirm the CDMS analysis that the DAMA claimed
modulation corresponds to a rate that should have been observed in CDMS\@.
The key question, however is, whether the same result applies for all
possible halo models, so we have calculated the expected rate and
modulation amplitude for CDMS/Edelweiss and DAMA as a function of WIMP mass
for a wide range of models.  To quantify our results, we present a novel
way of comparing models.  We consider the ratio of the rate from a
particular model to the rate from the standard model (for CDMS/Edelweiss),
and likewise the ratio of the modulation amplitudes for the models (for
DAMA).  If the ratio in the latter case is larger than 1, then the WIMP
cross section need not be as large to produce the same measured modulation.
This in turn will lower the best fit region from DAMA.  Alternatively, if
the former ratio is less than unity, this will raise the CDMS/Edelweiss
upper limit on the WIMP cross section.

\begin{figure*}
  \resizebox{5in}{!}{\rotatebox{-90}{\includegraphics{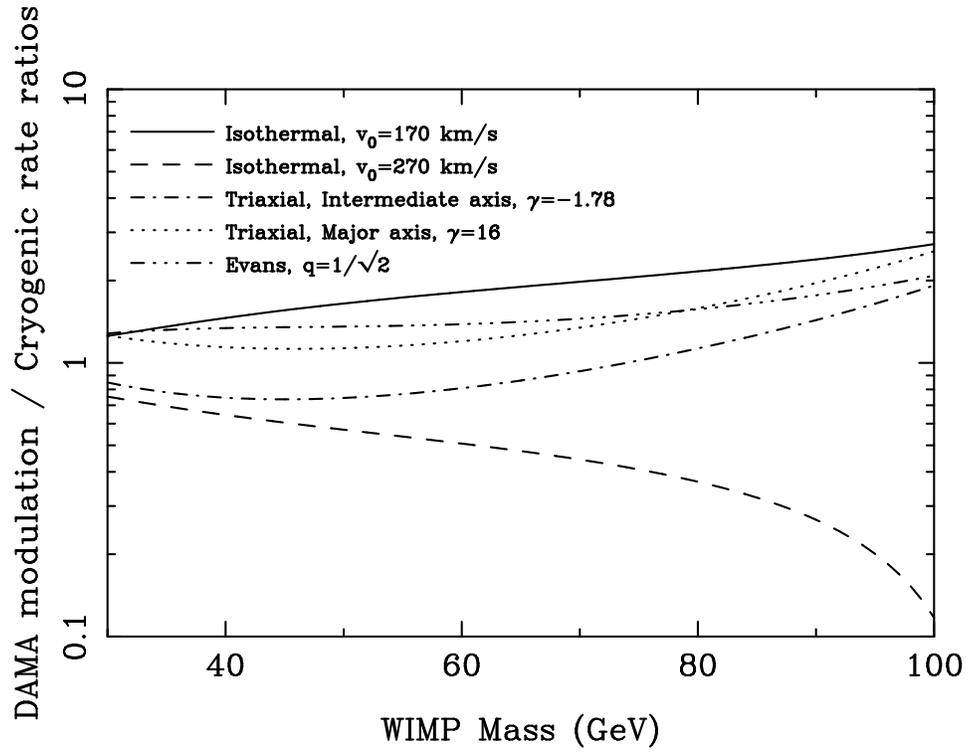}}}
  \caption{The ratio of the DAMA modulation ratio to the CDMS/Edelweiss
    rate ratio as a function of WIMP mass.  For the standard model this
    ratio is one.  Models for which this ratio is greater than one lead to
    better agreement between the experiments whereas models for which this
    ratio is less than one lead to more disagreement.}
  \label{fig:ratio}
\end{figure*}

Some variations will change both ratios in the same direction. For example,
decreasing the local halo WIMP density will lower the overall
CDMS/Edelweiss rate, however it will also lower the DAMA modulation
amplitude in the same way leading to no overall net effect in the
comparison of the two experiments.  We thus present a fiducial quantity
which cancels out such effects: we take the ratio of the DAMA modulation
ratio to the CDMS/Edelweiss rate ratio (figure~\ref{fig:ratio}) as a
function of mass.  For each mass the standard model is one.  Models for
which this ratio is greater than one lead to less overlap between the
CDMS/Edelweiss limit and the DAMA claimed observation, and hence reduce
disagreement between these experiments, whereas a ratio less than one leads
to more overlap and hence greater disagreement.

{}From figure~\ref{fig:ratio} one can see, as expected, that that a
narrower velocity dispersion in an isothermal distribution ($v_0=170\kmps$)
leads to a ratio greater than one and a broader isothermal distribution
($v_0=270\kmps$) leads to a ratio less than one.  In fact the narrower
isothermal distribution has the largest deviation from the standard model
in the direction necessary to reconcile the two experiments of any of the
models we examined. Most halo models fall between the two extremes set by
the isothermal models.  A flattened axisymmetric model (Evans model)
also provides some improvement over the standard model.  Triaxial models
can lead to either larger or smaller ratios for a detector with the
sensitivity of DAMA depending on the orientation of the axes.

Quantitatively, the maximum improvement we find for any model in the
agreement between CDMS/Edelweiss and DAMA is about 2.7 (at
$m_\chi=100\GeV$).  In general potential agreement between the experiments
improves at larger WIMP mass.  For the case of CDMS alone, this would open
up a potentially significant range of parameter space in which the DAMA
result might be compatible since CDMS is most sensitive at lower masses.
However, when the Edelweiss result is included, this factor would
effectively make the new Edelweiss limit equivalent to the older CDMS
standard halo model result at the higher masses, which is claimed to be incompatible with DAMA
as noted above.  Based on this claim then, the
CDMS/Edelweiss and DAMA are now seen to be incompatible, {\it independent
  of halo model uncertainties}.

There is one other important consideration in addition to the amplitude of
any annual modulation.  As seen from figure~\ref{fig:phase}, the phase of
the modulation in triaxial models will not in general agree with the
predicted phase from the standard isothermal model, which is in agreement
with the DAMA data.  This is an important effect.  Not only does it rule
out these models as possibilities for allowing a reconciliation of DAMA and
CDMS, if DAMA, or any eventual annual modulation observation has a phase in
agreement with the standard model, then these models will be severely
constrained.  Alternatively, and perhaps more interestingly, for low mass
WIMPs observing this phase of any annual modulation will tell us something
about the orientation of the axes in a triaxial model {\it without the need
  for a detector with angular sensitivity}.

As suggested by the above, the analysis performed here has application
beyond a mere comparison of DAMA and CDMS or Edelweiss, where our analysis
suggests the DAMA annual modulation is definitively in disagreement with
the latter two experiments.  The general method demonstrates both the
effect of galactic halo uncertainties on any comparison of overall
interaction rates with annual modulation in untangling WIMP parameters from
direct detection experiments, and also the possible utility of performing
such a comparison if eventually WIMPs are observed in both sets of
experiments.  In particular important information can be gleaned that might
compliment that which can be obtained in detectors with angular
sensitivity\cite{drift}.

\end{document}